# The threat of analytic flexibility in using large language models to simulate human data

Jamie Cummins
University of Bern, Switzerland
University of Oxford, United Kingdom

Social scientists are now using large language models to create "silicon samples": synthetic datasets intended to stand in for human respondents. However, producing these samples requires many analytic choices, including model selection, sampling parameters, prompt format, and the amount of demographic or contextual information provided. Across two studies, I examine whether these choices materially affect correspondence between silicon samples and human data. In Study 1, I generated 252 silicon-sample configurations for a controlled case study using two social-psychological scales, evaluating whether configurations recovered participant rankings, response distributions, and between-scale correlations. Configurations varied substantially across all three criteria, and configurations that performed well on one dimension often performed poorly on another. In Study 2, I extended this analysis to a published silicon-sample use case by re-examining Argyle et al.'s (2023) Study 3 using 66 alternative configurations. Correlations between human and silicon association structures differed substantially across configurations, from r = .23 to r = .84. Taken together, the results from these studies demonstrate that different defensible configuration choices can materially alter conclusions about the fidelity of silicon samples. I call for greater attention to the threat of analytic flexibility in using silicon samples and outline strategies that researchers may adopt to reduce this threat.

Generative large language models (LLMs) have ushered in a sea change within the social sciences. The capacity of these models to produce coherent natural language output to queries, as well as their flexibility, efficiency, and generalisability, has led researchers to explore their use in tasks across the research process. One of the most ambitious proposed uses of LLMs in this respect has been to use them as simulated research participants in order to determine how human participants might respond to a particular context, set of survey questions, or intervention (Z. Cui et al., 2025). This is based on the tacit assumption that, on some level, the corpora of data which LLMs are trained on may enable them to approximate similar data-generating processes to those which underpin human responses (Lin, 2025b). If achievable, this could have a transformative impact on the social sciences. It would enable researchers to predict the responses of vulnerable and/or hard-to-reach populations of participants without the need for time- and resource-intensive research studies, improving the representativeness and diversity of research (Cui et al., 2025). Researchers could iteratively pilot and improve their planned studies using LLMs prior to deploying them in human samples, improving the robustness of the research literature (Anthis et al., 2025; Säuberli et al., 2025). More generally, having access to LLMs which can act as valid surrogates for human participants could enable swifter comparison of multiple competing hypotheses (Zabaleta & Lehman, 2024). Indeed, it is difficult to overstate how revolutionary an effective "participant simulator" could be for the social sciences. The potential for such methods has not gone unnoticed: major companies such as Toluna, Ipsos, and Qualtrics are already marketing synthetic personas, synthetic participants, or synthetic panels as practical research tools which can be used as supplements to, or in place of, human respondents (Henriques, 2026; Ipsos, 2025; Toluna, 2026).

**Silicon samples: Promises and early applications**

Though only a recent phenomenon, researchers have already extensively used these silicon samples to simulate participants across the social sciences, including in psychology, economics, sociology, marketing, consumer behavior, and beyond. Indeed, results from some of these studies appear rather promising. Mei et al. (2024) for instance, found that both ChatGPT-3.5-Turbo and GPT-4 reproduced human-like behavior across six canonical games from economics, often aligning with average human choices. Similarly Marjieh et al. (2024) explored whether LLMs could be used to generate large-scale datasets of sensory judgments, finding that the similarity structures derived from these synthetic datasets aligned closely with human perceptual organization. On a larger scale, Cao et al. (2025) explicitly trained language models to approximate the distribution of responses from international survey datasets, showing that distribution-aware fine-tuning led to alignment between silicon samples and real survey responses. Further still, Dillion et al. (2023) highlighted strong correspondence between LLMs and humans on moral judgement tasks ($r > .90$), as well as LLM-human alignment in domains including voting behavior, problem-solving, and economic games.

The studies above generally looked at the capacity for LLMs to act as research participants without providing them with additional information about the type of participant they should act as. Other silicon sample research has additionally provided the model with specific demographic information (e.g., age, gender, income level) that it should account for when responding, varying the provided information across calls. For example, Hewitt et al. (2025) evaluated whether GPT-4 could simulate the results of social science experiments by creating a silicon sample of individual participants with specific demographic profiles. Using materials from 70 nationally representative survey experiments (476 treatment effects), they found that GPT-4-derived estimates of treatment effects were strongly correlated with the actual experimental outcomes ($r = 0.85$), and on average performed at least as well as human forecasters. In a similar vein, Saynova et al. (2025) tested whether LLMs could predict the replicability of behavioral social science studies, simulating participant responses with several LLMs based on 14 studies from the Many Labs 2 project. The authors concluded that LLMs could potentially be suitable for this purpose, with models achieving F1 scores as high as 77%. Complementary to this, others have constructed large silicon samples using GPT-3 by systematically varying sociodemographic variables included within prompts, showing that model outputs could approximate the distributions of human survey data across numerous psychological and political domains (Argyle et al., 2023). Qu & Wang (2024) used sociodemographic conditioning to simulate public opinion across World Values Survey items, reporting both promising aggregate alignments and systematic biases. Beyond academic contexts, Brand et al. (2023) demonstrated how LLM-generated silicon samples can be applied in marketing research, using sociodemographic prompts to model consumer preferences and thereby substituting for traditional survey samples.

**Limitations and emerging concerns**

The studies described above represent a small number of examples of work which seems to suggest silicon samples are feasible, useful, and accurate. However, not all work on the matter has been so positive in its assessment. Schröder et al. (2025) for instance, tested whether LLMs could reproduce human moral judgments when scenarios were subjected to subtle rewordings that preserved surface similarity but shifted semantic meaning. While human participants adjusted their evaluations to reflect the altered meaning, LLMs tended to ignore these differences and produced nearly identical ratings for both the original and reworded items. Lutz et al. (2025) reported a slightly more optimistic picture, showing that with careful prompt design and demographic alignment, some LLMs were able to approximate distributions of human responses across a range of social science tasks. However, they also demonstrated that even minor variations in prompt formatting could dramatically affect these outcomes. Indeed, LLMs' sensitivity to prompt features has been reported in other cases (e.g., Baumann et al., 2025), with authors noting that even seemingly-irrelevant prompt features such as formatting specifications can have a dramatic effect on output (Sclar et al., 2024). Even among those studies which concluded that LLMs could accurately simulate human responses, these are often accompanied with caveats (e.g., Hewitt et al. found that native predictions needed to be adjusted downwards as they tended to overestimate effects; Park et al. (2024) found less diversity among response distributions than in humans) or inconsistencies in results between studies (e.g., contrary to other findings, Suh et al. (2025) found fine-tuned LLMs could simulate response distributions accurately).

**Analytic flexibility in silicon samples**

Beyond general limitations in their implementation, there is one specific issue with the use of silicon samples that represents a critical threat: the substantial degree of analytic flexibility associated with their use. The ground of this garden of forking paths is well-trodden by social scientists; researchers have noted for years now that analytic flexibility can lead to variation in the results which are reported in scientific research (Gelman & Loken, 2013). In the context of p-value and null hypothesis significance testing, Simmons et al. (2011) demonstrated that undisclosed methodological and analytic flexibility can inflate false-positive rates to unexpectedly high values, which can in turn lead to subsequent issues with the replicability of associated inferences made based on these results (see also Stefan & Schönbrodt, 2023; Wicherts et al., 2016). These issues have been highlighted in existing fields of social science, including in fMRI methodologies (Carp, 2012) and large-scale survey research (Masur, 2023).

Research using large language models is equally, if not more, susceptible to these issues. There is necessarily a very wide range of analytic decisions that researchers must make when simulating human responses using LLMs, and these analytic decisions may have a major impact on output. Some of these issues have already reared their head in existing research. For example, different models can produce dramatically different results (Aher et al., 2022; D. Li et al., 2025; Xie et al., 2024). Other research has observed that model output is influenced by the specific demographic characteristics included in prompts (Sun et al., 2024), the presence or absence of fine-tuning models (Binz et al., 2025; Lu et al., 2025), and even the specific formatting of prompts (Sclar et al., 2024). At a more technical level, previous work has demonstrated that variations in the 'temperature' parameter when using LLMs produces substantial variation in silicon sample output (L. Li et al., 2025). Given the variability in output due to these researcher choices, this may contribute to the mixed conclusions in the literature on whether or not LLMs are good simulators of human responses.

There are further aspects of flexibility in silicon sample workflows that are, as yet, relatively unexamined. For instance, if having an LLM complete a series of different measures, should the items of these measures be presented (i) all-at-once in a single prompt, (ii) scale-by-scale, where separate calls are made for each scale, or (iii) item-by-item, where every item receives its own call per simulated participant? Separately, though temperature has received some degree of attention, other LLM hyperparameters such as top-k (which limits sampling to the k most likely tokens) or top-p (which samples from the smallest set of tokens whose cumulative probability exceeds p) have not. Further still, impact of changes in the "reasoning effort" parameter associated with several reasoning LLMs (such as OpenAI's ChatGPT-5 and the o-series of models) has also not yet been examined, nor has whether there is a difference between reasoning versus non-reasoning models generally. There are also unexamined questions relating to the degree of information provided to the model. For instance, whether the wording of items is provided identically to how they would be presented to participants, or whether supplementary information is also given (e.g., distributions of scores in the population, precise technical definitions of constructs). This can also be extended to how model output is treated; for instance, the scores returned by the model may be accepted as the to-be-analysed data as they are, or processed/rescaled in some manner to improve accuracy (Kambhatla et al., 2025). It is important to note, too, that many other degrees of freedom that are not mentioned here could plausibly be present in these workflows. Researchers have begun to make efforts to taxonomise the potential range of these decisions in different research contexts (Abdurahman et al., 2025; Lin, 2025a, 2025c), yet it is difficult to exhaustively document them, particularly when features of prompts can have a near infinite number of dimensions to vary. A nonexhaustive list of these examined and unexamined possibilities is provided in Table 1.

**Sleepwalking into the garden of forking paths**

Two things are now clear. First, the use of LLMs for the purpose of creating silicon samples is rapidly increasing across the social sciences. There is a clear hope and a growing belief that LLMs could be used to accurately simulate human responses to stimulus materials. If this is achievable, it would have profound implications for how research with human subjects is conducted. This is facilitated by the highly accessible nature of these models: commercial LLMs are general-purpose, low-cost, and highly intuitive (at least on the surface). Clearly, LLMs are already being used for this purpose, and will be used by more and more researchers for these purposes.

Second, silicon sampling is not a singular method: there are many decisions that need to be made in the process of generating a silicon sample and all of these analytic decisions, both individually and cumulatively, will influence the content of silicon samples - often substantially, and in ways that are not well-understood. This is a major threat to their utility and validity. Most critically, it means that if researchers have no good basis upon which to inform these methodological choices, then this analytic flexibility could potentially allow researchers to arrive at essentially any result using silicon samples, analogous to Simmons et al.'s (2011) warning that analytic flexibility in human subjects research "allows presenting anything as significant". Without identifying which parameters and decisions (if any) are most likely to render results consistent with human responses for a particular research question, the use of silicon samples is as likely to lead researchers astray as it is to guide them.

The combination of these two facts - the growing use of LLMs and the lack of good empirical justification for analytic decisions which affect output - means that the social sciences are currently sleepwalking into a literature filled with results grounded in arbitrary methodological decisions rather than principled and informed analytic choices. We already know the consequences of creating such a literature: unrobust and poorly-replicable findings, misled generations of researchers, and swathes of papers which ultimately provide little information about the domain they intended to explore. But given that silicon samples are still a relatively nascent tool, the social sciences have a rare opportunity to get ahead of this problem and set standards ahead of time for this literature. We can use the tools created, and lessons learned, from past mistakes to develop more principled standards and recommendations for when and how silicon samples are created, used, and evaluated within research.

**Table 1.** A list of some dimensions of analytic decisions that be made during the process of generating silicon samples, as well as some of the specific choices possible associated with these decisions.

| Dimension | Possible Analytic Decisions (non-exhaustive) |
|---|---|
| Model choice | - Chat assistant vs. reasoning model<br>- Specific model family (e.g., ChatGPT, Llama, Deepseek)<br>- Open-source vs. proprietary<br>- Fine-tuned vs. general purpose |
| Model hyperparameter settings | - Temperature specification<br>- Reasoning effort specification<br>- Top-k constraint<br>- Top-p constraint |
| Scale prompting strategy | - All-in-one<br>- Scale-by-scale<br>- Item-by-item |
| Demographics provided | - Age<br>- Gender<br>- Country of residence<br>- Country of origin<br>- Ethnicity<br>- Income level<br>- Political affiliation<br>- Performance on other tasks |
| Scale information and context provided | - Items only<br>- Definitions of scale constructs<br>- Psychometric norms of items |
| LLM calling approach | - Single call<br>- Multiple-calls-across-same-parameters<br>- Multiple-calls-across-different-parameters<br>- Multiple-calls-across-different-models<br>- Multiple-calls-across-different-parameters-and-models |
| Output handling and score postprocessing | - Native scores used<br>- Rescaled based on normative population distributions |

**This Research**

The first step in developing more principled standards for the use of silicon samples is to directly investigate the variation in silicon samples which can come about due to specific analytic decisions, and to determine whether there appears to be particular settings, parameters, or decisions that generally produce better silicon samples. To date, there has been no dedicated investigation into this question; studies which have seen effects of particular analytic decisions tend to only examine a single feature, while studies which have considered multiple features have tended to be conceptual, rather than empirical, in nature.

In this work, I conduct a direct and focused analysis of the impact of analytic flexibility on silicon sample output, with the aim of assessing whether even a small number of analytic decisions can affect consistency with observed human data. In Study 1, I use data consisting of 85 participants taken from a large-scale social and personality psychology dataset containing responses to psychological scales and extensive demographic information. I focus in particular on four analytic decision points mapped out in Table 1: (i) the model used, (ii) the temperature or reasoning effort selected, (iii) the degree of demographic information provided, and (iv) the scale sampling strategy. I evaluate LLM consistency with human data based on three different data features: accurately ranking participants, accurately estimating scale response distributions, and accurately estimating the correlation between two variables. These features in particular were chosen because they align with dimensions of Argyle et al.'s (2023) position on assessing algorithmic fidelity of silicon samples (specifically in terms of the Pattern Correspondence criterion) In Study 2, I extend this work further, and re-examine whether this analytic flexibility can have a material impact on the conclusions derived from an existing study using silicon samples (namely, Argyle et al., 2023, Study 3).

## Study 1

For Study 1, I examined how a small number of analytic choices may affect the output of silicon samples in terms of their performance relative to two commonly used social psychology scales. To do this, I used data from the Attitudes, Identities, and Individual Differences dataset (AIID; Hussey et al., 2022): a very large scale dataset of approximately 200,000 participants in total, each of whom completed a small subset of a wide array of measures from social and personality psychology, as well as a detailed set of demographic information questions. Measures were included in this dataset by the original researchers because they were topically relevant to social psychological research and are frequently used within such settings. I arbitrarily selected two scales from this dataset for examination: specifically, a preference measure of "gut feelings" towards African-Americans vs. European-Americans ("Gut Feelings Scale"; Ranganath et al., 2008) and the Belief In A Just World scale ("BJW Scale"; Hussey et al., 2022). The constructs behind both of these scales receive substantial attention in social psychology; for instance, the paper which originally conceived of the "belief in a just world" concept has been cited more than 1,800 times (Rubin & Peplau, 1975), while the assessment of gut feeling/"automatic" (racial) evaluations holds its own subfield within social psychology (e.g., Dovidio et al., 1997). Several studies in social psychology have also previously examined links between racial bias more generally and belief in a just world (e.g., da Costa Silva et al., 2018; Litam et al., 2022) As such, investigating the relationship between these two scales very much represents a plausible question that a social psychologist may have, and one they may wish to address using silicon samples.

### Method

#### Sample

In the AIID dataset, there were a total of 85 participants who provided complete data on both of these scales, as well as the demographic questionnaires.

#### Measures

**Belief In A Just World Scale.** The BJW scale is a 6-item measure which measures the belief that, in general, people get what they deserve. For each of the six items, participants respond on a Likert scale from 1 (strongly disagree) to 6 (strongly agree). The sum score from these items represents a global BJW score (Lipkus, 1991).

**Gut Feelings Scale.** The Gut Feelings scale consisted of a two-item scale which asked participants to separately rate gut feelings towards African-Americans and European-Americans on a scale from 1 (strongly negative) to 10 (strongly positive). A preference score is then calculated based on the difference in ratings towards the two groups. This scale, as well as similar such scales, are frequently used in social psychology to investigate preferences towards social groups (e.g., Ranganath et al., 2008).

**Demographic Questions.** The demographic questions asked of participants consisted of their age, gender, country of residence, education level, ethnicity, income level, and political identity. The specific wording of these items can be found in the codebook for the AIID dataset (https://osf.io/mxabf/).

#### Analytic Features and Silicon Samples

I selected four of the features mapped out in Table 1 to investigate the variation in silicon samples based on differing analytic decisions: the model used (gpt-5-mini-2025-08-07, gpt-5-nano-2025-08-07, gpt-4o-2024-08-06, gpt-4o-mini-2024-07-18, gpt-3.5-turbo-0125, o4-mini-2025-04-16, o3-mini-2025-01-31, llama-3.3-70b-versatile, and deepseek-r1-distill-llama-70b), the temperature or reasoning effort used (reasoning efforts of "high" and "low" applicable to 5-mini, 5-nano, o3-mini, and o4-mini; temperatures of 0, 0.5, 1, and 1.5 applicable to all other models), the degree of demographic information provided (none, age-and-gender only, or extensive information), and scale sampling strategy (all-in-one, scale-by-scale, or item-by-item). This led to a total of 252 possible combinations. For all of these 252 combinations, the LLM was required to generate predictions for each of 8 items (2 Gut Feelings items, 6 BJW items) for each of the 85 participants. After simulating all combinations, this led to a final silicon sample dataset of 21,420 simulated participants; after exclusions for refusals or other incomplete output, this left a total of 21,153 simulated participants with complete data. The Python code used to produce these samples is available at https://osf.io/skbqm/.

#### Analytic Strategy

Processing and analysis of the human and LLM data was conducted using R (R Core Team, 2025). All data, processing code, and analysis code is available at https://osf.io/skbqm/. Although there are many potential ways that we might analyse the accuracy of silicon samples relative to observed human data, here I focus on three specific data features for evaluation: ranking participants, estimating response distributions, and estimating the relationship between two scales. I also examine whether there is consistency among performance on each of these features.

In some cases, requests to the language models returned refusals to provide ratings, nonsense output (i.e., when temperature was set to 1.5), or text which was not in the prespecified format. In cases where valid ratings were given but text was not in the prespecified format, additional parsing was used to get the output into appropriate format for analysis. In cases where models refused to provide ratings or gave nonsensical output, all responses on the other scales for that participant generated within that LLM configuration were excluded; in other words, participants were only included for a given configuration if all responses returned a valid rating. In some cases, this led to a large quantity of incomplete data. Therefore, analyses were conducted only on configurations that returned complete data for at least 50% of all participants. Additionally, there were some cases where models gave identical numeric output for the entire sample (primarily when temperature was set to zero and no demographic information was provided). Consequently, correlations involving these scales could not be calculated (since the associated variance was zero) and therefore these observations were also excluded. These exclusions mean that estimates are conditional on the usable responses produced by each configuration. In particular, refusal or malformed output is part of configuration performance. Outputs with no variance should therefore be read as failures to generate usable variation for correlation-based criteria.

**Data Feature 1: Ranking Participants.** If a silicon sample has validly simulated participant responses, then the ordinal ranking for a given participant between the observed human data and the silicon sample should be roughly the same. In other words: if a participant's observed score is low, then the LLM should predict a relatively low score; if a participant's observed score is high, then the LLM should predict a relatively high score. The simplest way to examine this is to estimate the correlation between human scores and their corresponding LLM estimates for each configuration (and do this separately for each scale). The closer the correlation value to 1, the closer the ordinal ranking of participants is to the true ranking. For each configuration, I estimated Spearman's rho between participants' observed scale scores and the corresponding LLM-predicted scores, doing this separately for each scale. After exclusions, these correlations were conducted on 246 configurations for the BJW scale, and 233 configurations for the Gut Feelings scale. Importantly, readers should note that this is a test of pattern correspondence (Argyle et al., 2023), specifically in relation to the preservation of ordinal ranking among participants, and not individual-level prediction.

**Data Feature 2: Estimating Response Distributions.** Beyond the ordinal ranking of participants' responses to a particular scale, there remains an additional question: whether silicon samples accurately model the distribution of responses to that scale. Recall that the use of silicon samples comes with the assumption (and hope) that the data-generating process for the sample approximates that of the human sample. This does not, of course, mean that the same psychological processes are involved in the production of these data, but rather that the data themselves reflect how humans will behave or respond, irrespective of the data-generating processes at play (Lin, 2025c). By extension, if silicon samples are accurate, they ideally should yield response distributions which are similar to the human data.

To compare the correspondence of response distributions in silicon samples with human data, I used Wasserstein distance estimates (Kantorovich, 1960; Lutz et al., 2025). This metric quantifies the minimum effort required to transform one distribution into another by moving probability mass. In this sense, it provides a continuous measure of similarity between two distributions. I estimated Wasserstein distances between human data and data generated from each of the separate silicon sample configurations, doing this separately for both the BJW scale and the Gut Feelings scale. However, before estimating these distance values, I firstly scaled scores for each scale relative to their minimum and maximum bounds (i.e., -9 to 9 for the Gut Feelings scale; 6 to 36 for the BJW scale). This ensured that Wasserstein distance values were normalised between scales, allowing for a more interpretable comparison. I also estimated the "human" range of Wasserstein distances for each scale by (i) randomly splitting the human data into two halves, (ii) estimating Wasserstein distance between these two datasets, and (iii) repeating this process 2000 times. This provided a point estimate and split-half reference intervals for what the Wasserstein distance between two human samples for each scale may look like, thereby providing a baseline against which to compare the silicon samples' distances. After exclusions, Wasserstein distances were calculated for 251 configurations for both the BJW scale and the Gut Feelings scale.

**Data Feature 3: Estimating The Relationship Between Scales.** Beyond individual scale qualities, another dimension to evaluate the performance of silicon samples relates to how well they recover the relationship between two scales. After all, researchers are not only interested in the characteristics of individually-created variables, but also the relationships between them. If silicon samples provide accurate information about human responses, then the estimated correlation between two scales from the silicon sample should approximate that found in the corresponding human data. I therefore estimated the correlation between the BJW scale and Gut Feelings scale from each silicon sample configuration, and compared it to the ground-truth correlation between scales in the human data. After exclusions, this correlation was conducted on 232 pairs of BJW/Gut Feelings scale scores.

**Consistency Across Data Features.** In addition to examining the three above data features separately, I also sought to examine whether there was consistency in the performance of silicon sample configurations across these data features. In other words, if a silicon sample produced using a particular configuration performs well when evaluated based on one data feature (say, ranking participants), does this mean this sample is likely to be accurate for other data features too (say, estimating the relationship between scales)? To do this, I estimated the correlation between configurations' performances on the three features above, including the estimates for both scales focusing just on participants' rank-order and the sample distribution separately (leading to a total of five variables to be correlated: participant ranking for the BJW scale, participant ranking for the Gut Feelings scale, distribution approximation for the BJW scale, distribution approximation for the Gut Feelings scale, and the estimated relationship between variables). For the "participant ranking" scores, I used the same correlation value estimated in the Data Feature 1 analysis. For the "distributional approximation" scores, I used the Wasserstein distance estimates used in the Data Feature 2 analysis. For the "relationship between variables" scores, I used the absolute error between the predicted correlation for a given silicon sample and the ground-truth human correlation. For example, if the estimated correlation from a silicon sample was 0.4, and the correlation from the ground-truth human data was 0.65, then the absolute error score would be 0.25.

If a configuration's accuracy regarding one feature predicts its accuracy in terms of another, then we would expect that correlations between scores based on the first feature (ranking participants) and the score for the second feature (estimating the distribution of responses) to be negatively correlated (i.e., better ranking of participants should be associated with lower Wasserstein distance scores). Scores from ranking participants should additionally be negatively correlated with scores from the third feature (estimating correlation between scales; better ranking should be associated with lower absolute error in predicting scale correlations). Scores from the second and third features, by contrast, should be positively correlated, with lower scores in both implying better modelling of the data. Notably, since there are two scores for ranking participants (i.e., one for each scale), we would expect these to correlate highly with one another if configurations perform consistently between different scales. This can also be said for distributional approximation, given that we similarly have scores for each scale on this feature. For this comparison, configurations were only included if they produced complete data on all five scores from the three data features (i.e., ranking participants for BJW and Gut Feelings, response distributions for BJW and Gut Feelings, and the relationship between the scales). After exclusions, these correlations were estimated based on a total of 232 observations.

**Results**

Success completion rates from the configurations ranged from 0.21 to 1.00 but were generally very high ($M = 0.99$, median = 1.00). Only one configuration fell below the 50% threshold for the main analyses (o3-mini with low reasoning effort, scale-by-scale prompting, and minimal demographics). Several high-temperature DeepSeek-R1-distill-Llama-70b configurations also showed non-trivial attrition, primarily due to failures to conform to the required response output structure (not unexpected, given the high temperature setting). The specification-curve results reported below are only reported in reference to the usable outputs produced by the configurations.

**Data Feature 1: Ranking participants**

Figure 1 shows the specification curves for the correlation between participants scores in silicon samples and participant scores in observed human data for both the BJW scale and the Gut Feelings scale. The observed range of correlations was $-0.24 < \rho < 0.40$ for the BJW scale and $-0.29 < \rho < 0.22$ for the Gut Feelings scale, indicating that there was both substantial variation in estimated correlations, but none rising beyond a relatively low value. Although there did not appear to be a clear pattern for configurations in the data, correlations generally were greater (although still rather small) when extensive demographics were used (vs. minimal or none).

**Data Feature 2: Estimating Response Distributions**

The observed Wasserstein distances for the data from the LLM configurations for both scales can be seen in Figure 2. The estimated Wasserstein distance values between halves of the human data were $W = 0.05$, 95% split-half reference interval [0.02, 0.09] for the BJW scale, and $W = 0.03$, 95% split-half reference interval [0.01, 0.05] for the Gut Feelings scale. By comparison, equivalent Wasserstein distance means and central 95% specification ranges across LLM configurations were measurably larger; BJW scale mean $W = 0.19$, central 95% specification range [0.04, 0.36]; Gut Feelings scale mean $W = 0.07$, central 95% specification range [0.03, 0.12].

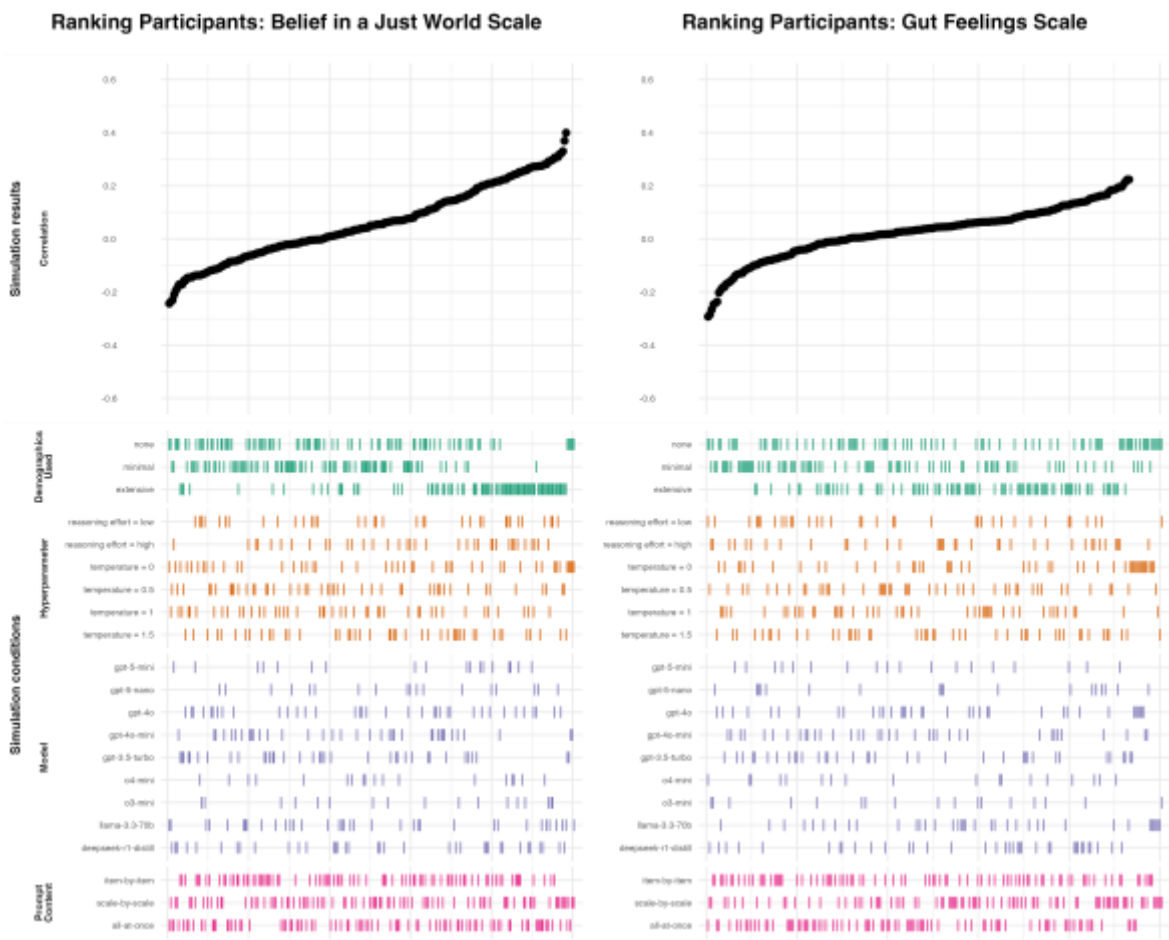

**Figure 1.** Specification curve for estimating human-LLM correlations between predicted participant scores, separately for the Belief in a Just World and Gut Feelings scales towards European-Americans vs. African-Americans items.

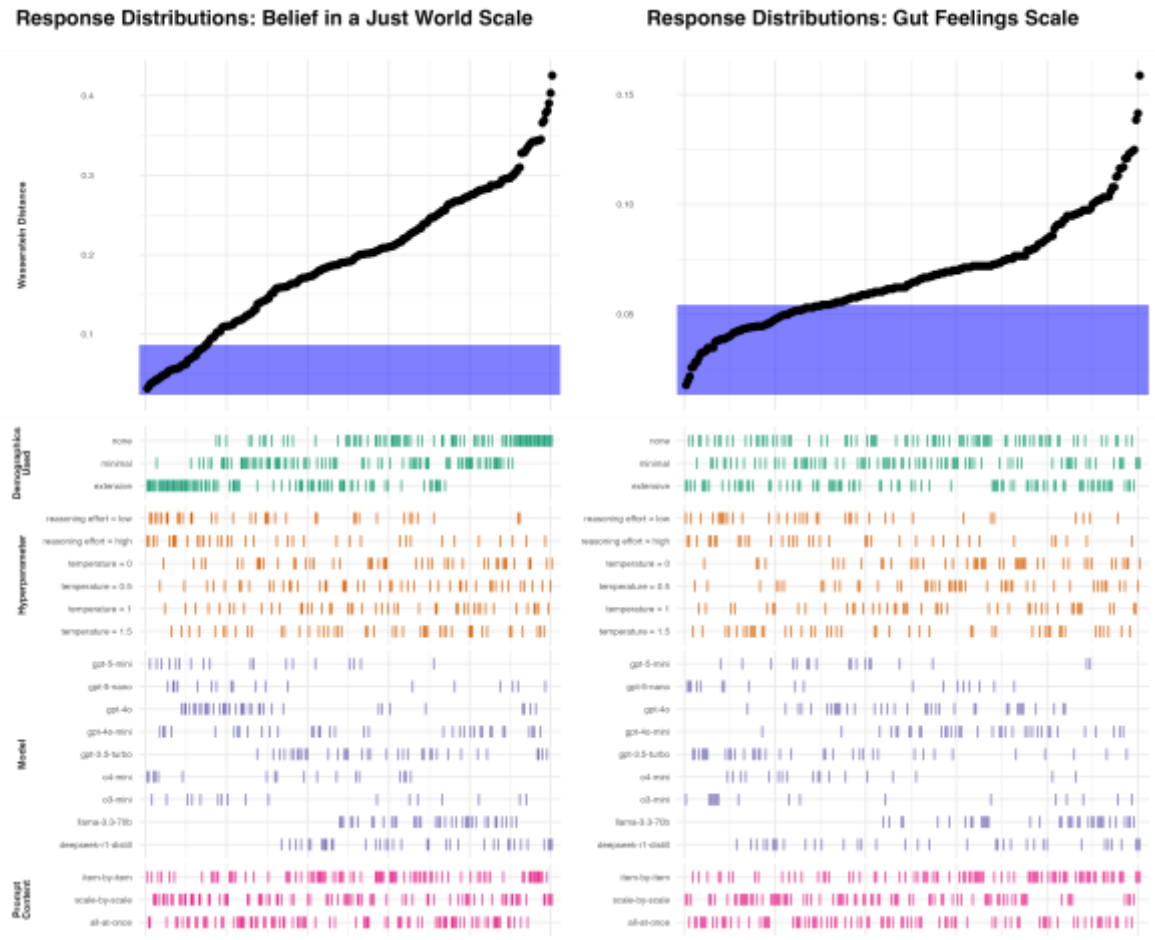

**Figure 2.** Specification curve for the observed Wasserstein distances for each configuration in both scales. The blue-coloured region in each upper panel represents the split-half reference interval of Wasserstein distances observed in the human-human distance comparisons for that scale.

**Data Feature 3: Estimating the correlation between variables**

Figure 3 presents these estimated correlations between the BJW and Gut Feelings scales based on each of the silicon sample configurations. The observed empirical correlation between the BJW scale and the Gut Feelings scale in the observed human data was $r = 0.26$, 95% CI [0.05, 0.45]. This estimated correlation in the data from the different silicon sample configurations varied dramatically (range: $-0.40 < \hat{r} < 0.78$), though notably the overall average point estimate across all configurations ($\hat{r} = 0.21$) was a close approximation of the true point estimate. Notably, although some configurations estimated the correlation very closely, minor changes affected predictions dramatically. For instance, the most accurate prediction ($\hat{r}$ = 0.24) was produced using ChatGPT-3.5 with a temperature set to 1, extensive demographic information provided, and presentation of scale items item-by-item. However, taking those same settings but increasing temperature to 1.5 made the model underestimate the correlation by 0.44 ($\hat{r}$ = -0.18); decreasing temperature to 0.5 also made the model drastically underestimate the correlation by 0.26 ($\hat{r}$ = 0.00), while decreasing temperature to 0 led the model to only slightly underestimate the correlation ($\hat{r}$ = 0.17).

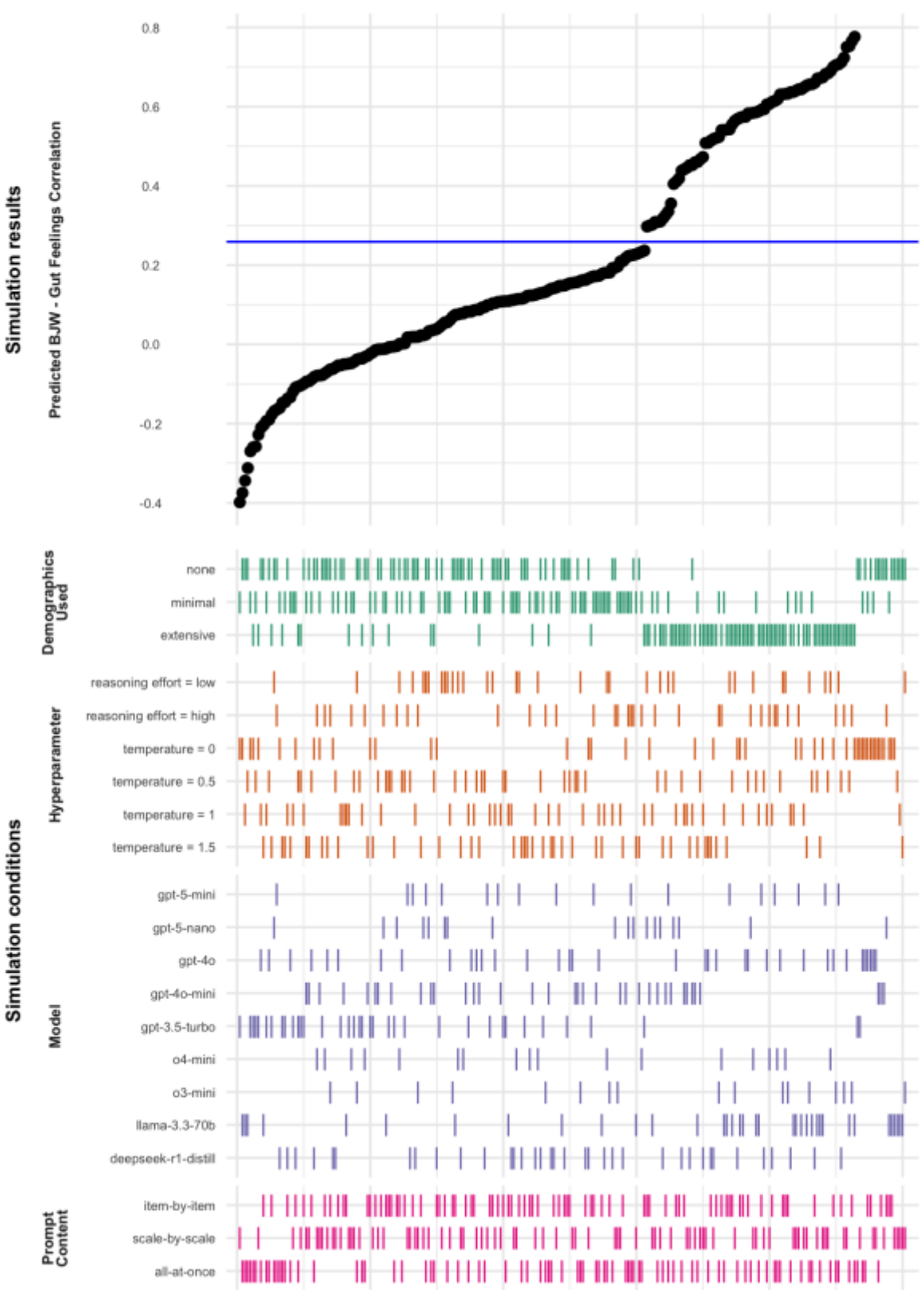

**Figure 3.** Specification curve for the estimate of the correlation between the BJW scale and the Gut Feelings scale. The blue line represents the point estimate for the empirical correlation between the BJW scale and the Gut Feelings scale observed in the human data.

### Consistency Across Data Features

The correlations between the data features are presented below in Table 2. Notably, the correlations were varied, with most being rather small in magnitude. Even if a configuration for a silicon sample workflow was found to be highly similar to human data in terms of one data feature, this does not mean that it also possessed similar levels of quality for other data features. Interestingly, for distributional approximation, there appeared to be no significant correlation between the two scales: in other words, there was no significant relationship between how well the same silicon sample configuration modelled the distribution of one scale vs. another. On the other hand, the performance of a configuration in predicting participant ranking in one scale was significantly correlated with its predictions of participant ranking in the other scale. Perhaps most surprisingly, performance in ranking participants (for both scales) was significantly positively correlated with absolute error in predicting the relationship between scales; this implies that better modelling of scale responses was associated with greater error in predicting the relationship between scales.

**Table 2.** Correlations between configurations' performances on each of the data features and each of the scales.

| Variable | 1 | 2 | 3 | 4 |
|---|---|---|---|---|
| 1. Data Feature 1: BJW | | | | |
| 2. Data Feature 1: Gut Feelings | .25** [.12, .37] | | | |
| 3. Data Feature 2: BJW | -.39** [-.50, -.28] | -.08 [-.21, .05] | | |
| 4. Data Feature 2: Gut Feelings | .08 [-.05, .21] | -.04 [-.17, .08] | .12 [-.01, .25] | |
| 5. Data Feature 3 | .24** [.12, .36] | .10 [-.03, .23] | -.13* [-.25, -.00] | .07 [-.06, .19] |

*Note.* M and SD are used to represent mean and standard deviation, respectively. Values in square brackets indicate the 95% confidence interval for each correlation. The confidence interval is a plausible range of population correlations that could have caused the sample correlation (Cumming, 2014). * indicates $p < .05$. ** indicates $p < .01$.

### Exploratory Additions

**Configuration Features Predictive of Performance.** At the request of a reviewer, I ran a set of exploratory univariate regressions to determine whether any of (i) model type (reasoning vs. non-reasoning), (ii) demographics detail, and (iii) prompt content tended on average to predict better performance on any of the evaluation dimensions. For these regressions, all outcomes were recoded so that higher scores = superior predictive performance. Additionally, because multiple comparisons for each predictor were made (i.e., five different evaluation feature elements), a Bonferroni-corrected alpha was used (i.e., 0.01). Each model used one predictor family at a time (model family, demographics used, or prompt content) and one outcome at a time. Results are presented in Table 3. In general, model family predicted the two distributional-fit outcomes, while demographic information predicted between-scale relationship accuracy, BJW distributional fit, and both ranking outcomes. Prompt-content omnibus effects significantly predicted Gut Feelings distributional fit only.

**Accounting for Sampling Error.** A fair criticism of the above would be that the observed variability may not exceed that which would be expected from human samples themselves. I explored this question in the context of recovering the human BJW-Gut Feelings between-scale correlation. Specifically, I repeatedly split the human data into two halves and examined the correlation in each half. Across 1,000 repeated non-overlapping human splits, the independent-human benchmark error had a median of $|\Delta r| = .24$, with a 95% interval from .01 to .70. Across all silicon configuration-split combinations, the corresponding silicon error had a median of $|\Delta r| = .26$, with a 95% interval from .01 to .73. On the one hand, this implies that silicon samples performed relatively well. Indeed, because the configuration space is large, at least one silicon configuration beat the independent-human benchmark in 99.7% of splits, even though the mean share of configurations doing so within a split was only 45.4%.

Importantly, however, variability across splits in terms of specific configuration performance was substantial. The independent-human benchmark beat the median configuration in 55% of evaluable splits. The best-performing silicon sample configuration, interestingly, beat the human sample in 89% of its evaluable splits. On the surface, this may imply that this configuration may be useful. However, this configuration consisted of the ChatGPT-nano model on high reasoning effort, scale-by-scale prompting, and no demographic information. This runs contra to the idea of a silicon sample, which conditions its participants on specific demographic information. Indeed, the 7 best-performing silicon samples (vs. the human data) contained at most minimal demographic information.

I also examined whether the observed specification spread could plausibly be attributed mainly to stochastic variability within a fixed configuration (cf. J. Cui & Alexander, 2026). To do so, I examined only those configurations which involved a temperature setting at zero, which would minimise the degree of variability observed across calls. Of the 31 analysable configurations with this temperature setting, the observed range of correlations was -.40 to .78 (SD = .37). Although this analysis does not estimate within-configuration variance directly, it does show that substantial spread can be observed even in a low-temperature subset.

**Table 3.** Omnibus tests of whether configuration features predict Study 1 performance across evaluation outcomes. For these exploratory models only, all outcomes are oriented such that higher values indicate better performance. Bonferroni-adjusted significance is evaluated within each predictor family across the five outcome comparisons.

| Predictor | Outcome | F(df1, df2) | p |
|---|---|---|---|
| Model family | Between-scale relationship accuracy | F(1, 230) = 1.542 | .216 |
| Model family | Distribution fit (BJW) | F(1, 249) = 33.237 | < .001* |
| Model family | Distribution fit (Gut Feelings) | F(1, 249) = 22.592 | < .001* |
| Model family | Ranking accuracy (BJW) | F(1, 244) = 5.832 | .016 |
| Model family | Ranking accuracy (Gut Feelings) | F(1, 231) = 0.415 | .520 |
| Demographics used | Between-scale relationship accuracy | F(2, 229) = 9.790 | < .001* |
| Demographics used | Distribution fit (BJW) | F(2, 248) = 64.125 | < .001* |
| Demographics used | Distribution fit (Gut Feelings) | F(2, 248) = 0.259 | .772 |
| Demographics used | Ranking accuracy (BJW) | F(2, 243) = 70.727 | < .001* |
| Demographics used | Ranking accuracy (Gut Feelings) | F(2, 230) = 14.606 | < .001* |
| Prompt content | Between-scale relationship accuracy | F(2, 229) = 1.455 | .236 |
| Prompt content | Distribution fit (BJW) | F(2, 248) = 1.485 | .228 |
| Prompt content | Distribution fit (Gut Feelings) | F(2, 248) = 9.566 | <.001* |
| Prompt content | Ranking accuracy (BJW) | F(2, 243) = 0.317 | .728 |
| Prompt content | Ranking accuracy (Gut Feelings) | F(2, 230) = 3.277 | .040 |

**Discussion**

The results of the above study suggest that there is substantial variability to be found in silicon samples as a function of only a small number of analytic decisions that can be made. There is no apparent orderliness to these analytic decisions, in the sense that performance of configurations is not uniform nor consistent across different data features: a configuration may excel in modelling one feature while failing entirely in another. The variation observed here can also be seen when using one half of the human data to predict correlations in the other half, and silicon samples often performed similarly to (or even exceeded) the human splits. In cases where specific configurations most frequently beat the human data across these splits, these largely consisted of configurations that used little (if any) demographic information – running contra to the premise of silicon samples in the first place.

## Study 2

Although Study 1's results are striking, it is not clear the extent to which they may represent an isolated observation specific to the two social-psychological scales chosen, or whether the observed flexibility may influence the outcome of existing research findings which used silicon samples. Study 2 aimed to examine exactly this: whether applying the flexibility quantification used in Study 1 may reveal hidden effects of flexibility in already-published work. One of the most highly cited papers using silicon samples is Argyle et al.'s flagship 2023 paper, which has been cited more than 1300 times since its publication in 2023. In this work, the authors not only proposed an agenda for evaluating the algorithmic fidelity of silicon samples (which inspired the evaluation approach used in Study 1) but also demonstrated the apparent utility of silicon samples across three studies. Their third study examined whether silicon samples could replicate patterns of association among multiple demographic, attitudinal, and political variables, conditioned on other observed variables from the same respondents. More specifically, using human data from the 2016 ANES survey (American National Election Studies et al., 2017), the authors used twelve interview variables (with a vote choice variable conditional on a reported turnout variable), to produce eleven analysed items. Across calls to an LLM, they used an interview-style response template to solicit a predicted response for a held-out target item conditioned on the respondent's observed values for the remaining available variables, and observed relatively high agreement between the silicon sample predictions and the observations in the human data. In Study 2, I repeated this procedure using the same general data structure, while examining whether different silicon sample configurations would lead to different estimated silicon-human agreement.

## Method

### Sample

The human reference data came from the 2016 ANES Time Series Study (American National Election Studies et al., 2017), following the setup used in Argyle et al.'s Study 3 with one key change: to reduce costs, I used only the first 500 complete-case respondents, rather than using the full set of approximately 2,400 respondents. The full sample consisted of respondent data for whom the twelve interview variables used in the replication task could be represented in the prompting and evaluation data. Because vote choice was conditional on having voted, these twelve interview variables corresponded to eleven unique analysed items. Each respondent therefore served as a target case for the silicon-sample procedure: for a given analysed target item, the model was prompted with that respondent's observed values on the remaining available variables and asked to generate the held-out response. The same 500 human respondents were used across all silicon-sampling configurations.

### Measures

Following Argyle et al. (2023), Study 2 used the twelve 2016 ANES variables included in their Study 3 interview-style task: gender, race/ethnicity, age, education, church attendance, feelings of patriotism when seeing the American flag, whether the respondent discusses politics with family or friends, political interest, 2016 vote turnout, 2016 presidential vote choice, 7-point ideological self-placement, and 7-point party identification. These variables span demographic, religious, attitudinal, political, and behavioral information. As in Argyle et al.'s original analysis, vote turnout and vote choice are linked because vote choice is conditional on having voted; consequently, the twelve interview variables correspond to eleven unique analysed items. Full question wording, coding details, and template text are provided in Argyle et al. (2023).

### Analytic Features of Silicon Samples

Similar to Study 1, I examined four potential sources of analytic flexibility: model choice, temperature/reasoning effort, prompt content, and demographic information provided. For model choice, I examined gpt-5-mini, gpt-4o-mini, gpt-3.5-turbo, and llama-3.3. For temperature/reasoning effort, I examined low and high reasoning effort, and temperature values of 0, 0.7 (the main temperature used in Argyle et al.) and 1.5. For prompt content, I used one of three approaches: Argyle et al.'s original "interview-style" prompt, or an alternative "narrative profile" style prompt (similar to that used in other research, such as Park et al., 2024), or a "bullet point" approach which simply listed characteristics as bullet points before eliciting a response. For conditioning information, I provided either the full set of remaining available variables for each target item, or a reduced "demographics plus engagement" profile. The reduced profile included gender, race/ethnicity, age, education, church attendance, whether respondents discuss politics with family and friends, and political interest, while omitting the remaining attitudinal and explicitly political variables. This resulted in a total of 66 configurations for comparison.

### Analytic Strategy

For each configuration, I evaluated whether the resulting silicon sample reproduced the pattern-correspondence result reported by Argyle et al. (2023). Following their analytic approach, I computed Cramer's V for each pairwise association among the analysed ANES items in the human data. I then computed the corresponding Cramer's V values in the silicon data, pairing the model-generated response for a held-out target variable with the observed human values on the conditioning variables. This produced, for each configuration, a vector of silicon-sample associations that could be compared directly against the corresponding vector of human-sample associations. This is, in essence, the same analytic approach originally used by Argyle et al.

The primary variable of interest was the correlation between the vector of human Cramer's V values and the corresponding vector of silicon Cramer's V values. Higher values would indicate closer recovery of the rank-order pattern of associations in the human data, and therefore provide a direct test of the pattern-correspondence criterion emphasized by Argyle et al. (2023). As a secondary summary, I also computed the mean absolute difference between the human and silicon Cramer's V values across these pairwise associations, with smaller values indicating closer recovery of the absolute magnitude of the human association structure. I then examined the range of configuration-level performance across the 66 silicon-sampling specifications. This allowed me to ask not only whether the general pattern reported by Argyle et al. could be recovered, but the extent to which this recovery depended on specific silicon sample configurations.

## Results

Across the 66 silicon-sampling configurations, the estimated association structure among items varied substantially. The original analysis conducted by Argyle et al. required valid generated responses across all eleven analysed items. Not every attempted configuration produced an analysable matrix-level estimate, because some configurations led to refusals to produce gender, age, or race predictions. In total, 53 of the 66 configurations produced sufficient usable output for the all-item specification curve. Among these analyzable configurations, the correlation between the vector of human Cramer's V values and the corresponding vector of silicon Cramer's V values ranged from $r = .23$ to $r = .84$, with a median of $r = .58$ (correlation observed in original data: $r = .72$). In other words: the same human dataset using the same general Argyle-style prediction task could imply either relatively strong or weak pattern correspondence depending on the configuration used. For reference, I also recomputed the performance of Argyle et al.'s original generated silicon sample. The GPT-3 matrix correlated $r = .72$ with the human matrix, with a mean absolute difference of .05. The results of the flexibility analysis are displayed in Figure 4, with a reference line added for Argyle et al.'s original findings.

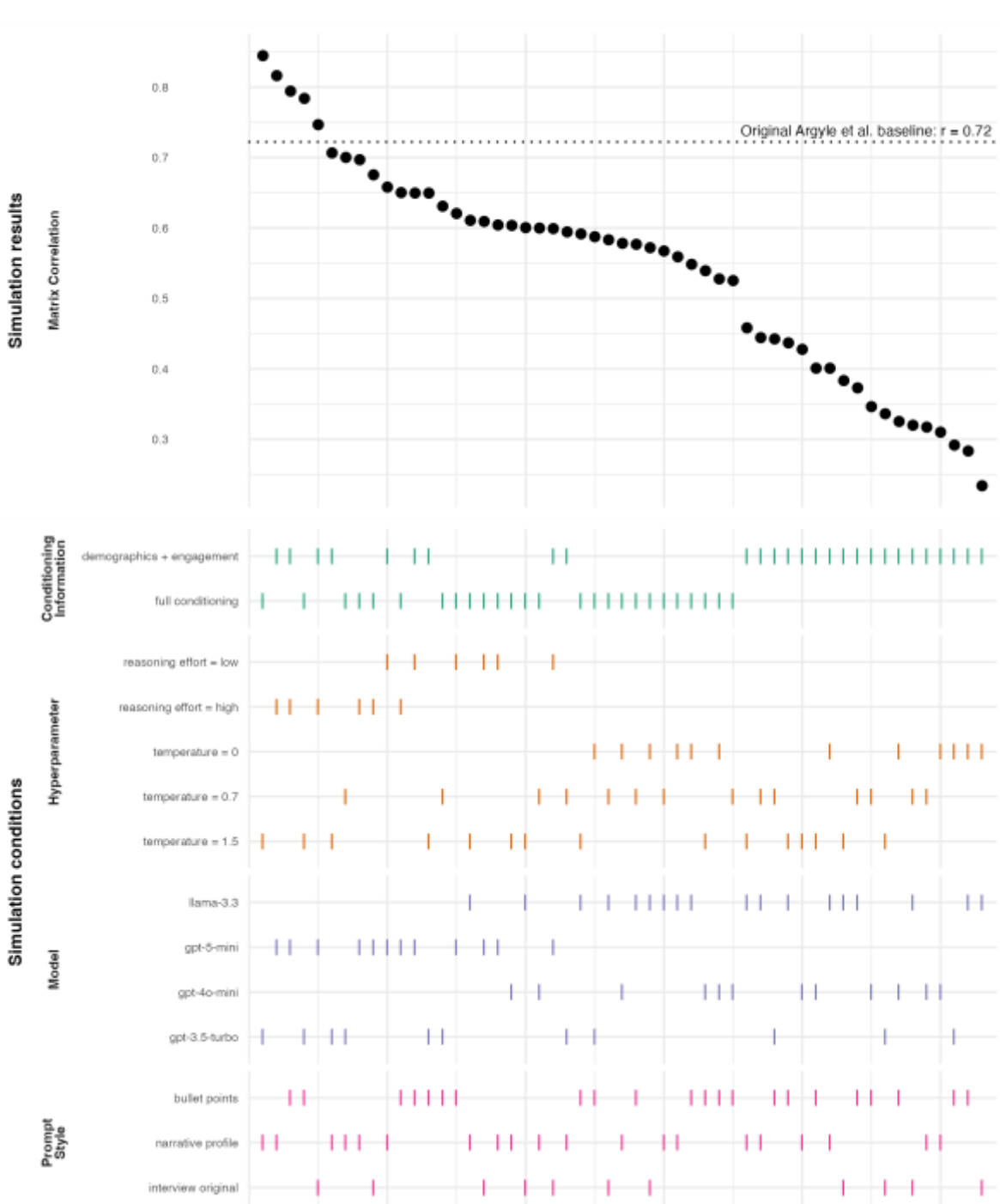

**Figure 4.** Specification curve for Study 2. Points show configuration-level correlations between the human Cramer's V association matrix and the corresponding silicon-sample Cramer's V association matrix across all analysable all-item configurations.

The mean absolute difference between human and silicon Cramer's V values showed a similar pattern. Across analysable all-item configurations, the mean absolute difference ranged from .08 to .19, with a median of .12 (error observed in original data: mean absolute difference = .05).

The all-item analysis also revealed substantial variation in usable output. Across all 66 attempted configurations, the number of respondents with complete generated data across the eleven analysed items ranged from 0 to 500, with a median of 338. Among configurations that were analysable, the complete-case sample size ranged from 6 to 500, with a median of 424. This means that the specification curve is itself conditioned on configurations producing usable output, a point that indeed reinforces the concern about flexibility in silicon-sample workflows. Notably, the spread was not limited to the lowest-N configurations: among configurations with at least 100 complete cases, matrix correlations still ranged from r = .23 to r = .78.

The identity of the best-performing configurations also depended on the analytic choices varied in the specification grid. In the all-item analysis, the highest-performing configuration used gpt-3.5-turbo at temperature = 1.5 with a narrative-profile prompt and full conditioning information, producing r = .84 and mean absolute difference = .09. By contrast, the lowest-performing analyzable configuration used llama-3.3 at temperature = 0 with the original interview-style prompt and the reduced demographics-plus-engagement profile, producing r = .23 and mean absolute difference = .15. More generally, the ten best-performing all-item configurations were not simply the original Argyle-style prompt or the richest-information condition: two used the original interview-style prompt, six used the narrative-profile prompt, and two used the bullet-point prompt; five used full conditioning information and five used the reduced demographics-plus-engagement profile.

To check whether the all-item result was driven by failures on age, gender, or race, I also repeated the analysis but excluding those three target variables. This constrained analysis retained all 66 configurations and again showed substantial flexibility. Correlations between the human and silicon association structures ranged from r = .01 to r = .86, with a median of r = .64, while mean absolute differences ranged from .05 to .17, with a median of .12.

## Discussion

Study 2 extended the Study 1 results to a direct reanalysis of a highly cited published silicon-sample demonstration (namely, Study 3 of Argyle et al., 2023). The consistency of the silicon samples in predicting human responses varied substantially across configurations. The best-performing configurations suggested relatively good recovery of the human association matrix, but other plausible configurations produced much weaker correspondence (or failed to produce analysable all-item matrices at all). A researcher using only one configuration could thus reach a much more optimistic or pessimistic conclusion about silicon sample fidelity than a researcher using another, even in the context of existing published work.

## General Discussion

Across two studies, I found that silicon-sample outputs were highly sensitive to analytic choices made during sample generation. Study 1 demonstrated this using two commonly used social-psychological scales. Decisions on just four choice criteria (model, temperature/reasoning effort, prompt format, and demographic information provided) led to substantial variation in the quality of silicon samples' recovery of participant rankings, response distributions, and between-scale correlations. Critically, there did not appear to be a discrete one-size-fits-all set of decisions: the performance of configurations was generally not correlated across different data features.

Study 2 extended this line of reasoning and examined the extent to which flexibility impacts results of a published, highly-cited use of silicon samples. Re-examining Argyle et al.'s (2023) Study 3, I found that the apparent recovery of the human association structure depended substantially on the silicon-sampling configuration used: correlations between human and silicon Cramer's V association structures ranged from r = .23 to r = .84, while mean absolute differences ranged from .08 to .19. Several configurations were not analysable due to blanket refusals to produce gender, age, or race predictions: when excluding these three variables from analysis and analysing all configurations, correlations ranged from r = .01 to r = .86. These results highlight that Study 1's findings extend beyond a specific demonstration in social-psychological scales; analytic flexibility can undermine even existing published silicon sample research findings.

## Implications for Silicon Samples

The analytic choices made in creating silicon samples can produce dramatically different estimates for even simple effects of interest – both in a case study (Study 1) and in data used in published silicon sampling studies (Study 2). Moreover, the idea that we could simply find the "best" configuration for creating silicon samples, even when considering only a small subset of the total range of decisions, does not appear to be feasible - the correlation in the accuracy of configurations across different data features appears quite small at best (cf. Study 1).

Notably, most (if not all) of the analytic choices deployed in both studies are in principle defensible. Study 2 in particular demonstrates that such choices can dramatically affect conclusions of silicon sample fidelity even in the case of published work. This has strong implications for claims relating to silicon samples as a method in general: findings of the fidelity of a silicon sampling approach in a given assessment context, with a given configuration, cannot readily be generalised beyond those contexts and configurations.

Silicon samples are increasingly being discussed and explored for their potential as practical tools for study planning, pilot testing, sample replacement, or conducting research with hard-to-reach populations. If researchers are to use silicon samples to make these decisions, they might reasonably expect that the decisions they arrive at are based on true signals from data, rather than artefactual effects of analytic choices. If not, the downstream scientific decisions or conclusions that researchers make will be largely driven by configuration-induced variation (likely without the awareness of the decision-maker) rather than by the substantive effects of interest.

One important condition of these results is that both studies concern survey-style persona prompting: models were asked to generate responses to structured survey items conditional on demographic information. These results should therefore not be assumed to generalise automatically to synthetic data settings writ large, particularly in the context of agent-based behavioural simulations. Configuration choices may matter in those settings as well, but the relevant degrees of freedom are likely to differ from those examined here.

Notably, it appears that researchers cannot simply look to pre-existing work to determine the best configuration of interest. As demonstrated in Study 1, the performance of different configurations showed nonuniform patterns of correlation between data features; in some cases, better performance on one data feature was associated with worse performance on another data feature. It appears, further, that configurations may vary in performance across the substantive domains in which they are tested. For example, in Study 1, GPT-3.5 Turbo with full demographic information and temperature at 1.5 performed relatively poorly when evaluated on recovery of the BJW-Gut Feelings association, with a median absolute relationship error of .406 across prompt formats, placing it at the 27th percentile of configuration performers. In Study 2, by contrast, this configuration was highest performer for recovery of the Cramer's V association matrix, with a median matrix correlation of .814 across prompt formats. Put simply, this implies that the performance of a given configuration cannot be used to generalise beyond (i) the specific domain, and (ii) the specific data feature examined.

If specific analytic choices can dramatically alter silicon sample performance, and observed silicon sample performance cannot generalise beyond the domain and data feature they are observed in, then this is a substantial threat to the use of silicon samples in research. If researchers wish to continue using these methods, then they must employ more rigorous methods of development to ensure that the analytic decisions they have opted for are sufficiently performant in the specific context of interest. In the next section, I attempt to offer some practical guidance on how such development might be undertaken.

## Practical Guidance for Silicon-Sample Workflows

To reiterate the foregoing point: the most immediate recommendation arising from the results of this work is that researchers should treat a silicon sample as the product of a well-defined development workflow focused on the specific domain(s) and data feature(s) of interest. Before generating a silicon sample, researchers should define (and, ideally, preregister) the intended use of the sample, the human benchmark against which performance will be evaluated, and a criterion against which to evaluate performance (i.e., a specific data feature to be modelled). Without specifying the target criterion in advance, researchers risk selecting configurations that appear successful only after seeing the result.

Second, researchers should consider and map plausible analytic decisions and configurations before generating and analysing silicon samples. Ideally, researchers should further use specification curves (or equivalent multiverse summaries) to document how their conclusions vary across them. This should be done to make plausible researcher degrees of freedom visible to readers, which can help readers both evaluate the robustness of results, and potentially provide valuable information for future researchers. Where possible, this specification set should also be preregistered before examining the focal outcome. The Study 2 results make this recommendation concrete. If the analysis had considered only the best-performing configurations, one could conclude that the Argyle-style association-recovery task was largely successful. If it had considered only weaker nearby configurations, the conclusion would be much more skeptical. A specification curve makes the ambiguity visible.

Third, researchers should explicitly separate steps of calibration and confirmation in the development of silicon samples, in a manner analogous to norms within machine learning research (i.e., train-test splitting). If a human benchmark dataset is available, one such workflow may involve splitting this benchmark data into calibration and confirmation subsets. The calibration subset may be used to compare plausible silicon-sampling configurations and to select a best-performing configuration using the prespecified quality criterion of interest (e.g., modelling response distributions). The selected configuration may then be used to generate a new silicon sample based on the confirmation subset, and its performance in this held out subset may then give insight into the potential validity of this configuration to data more generally.

I briefly explored the plausibility and limitations of such a calibration-confirmation approach using the data from both studies reported here, focusing on estimating between-scale correlation in Study 1, and overall Cramer's V correlation in Study 2. In both cases, I randomly split the human participants into two sets of equal size: a calibration set and a confirmation set. In the calibration set, I identified the silicon-sampling configuration whose estimated outcome (Study 1: BJW-Gut Feelings correlation; Study 2: Cramer's V matrix) most closely matched the corresponding human values. For Study 1, the best performing configuration (o4-mini, low reasoning effort, scale-by-scale prompting, no demographics) exhibited a calibration-set correlation of $r = 0.169$ compared with the human calibration-set correlation of $r = 0.165$ (absolute error = 0.003). I then evaluated that same configuration using the confirmation set. The estimated silicon correlation for this set was $r = -0.02$, compared to the true observed correlation of $r = 0.357$ (absolute error = 0.381). For Study 2, the best performing configuration in the calibration set (GPT-3.5 Turbo, temperature 1.5, narrative profile prompting, full demographic profile) exhibited an estimated correlation with the human Cramer's V matrix of $r = 0.877$. When applying this configuration in the confirmation set, the estimated correlation for this set was $r = 0.804$.

The discrepancy between the quality of this calibration-confirmation approach between the studies can likely be attributed to differences in sample sizes: given the relatively lower sample size of Study 1, there is necessarily greater noise and uncertainty in the estimated correlations. On the other hand, other explanations are also possible, including instability of the correlation metric in small samples, overfitting to a noisy calibration split, and mismatch between the calibration criterion and the broader task. However, regardless of the specific reasons, this discrepancy further emphasises the core point that the rationale behind specific silicon sample configuration decisions must be given extensive attention, including the acquisition of a sufficiently large calibration sample to ensure robust estimates of the criterion of interest.

**Practical Questions to Consider Before Using Silicon Samples**

Given the current findings, the final recommendation for researchers is to consider in the first place whether silicon samples are a useful, viable, or appropriate approach for their use case. For instance, consider the question of estimating correlations between two or several scales. One could, in principle, use silicon samples to approach this question. However, other methods may in fact be more appropriate. Specifically, researchers could instead use the relative semantic similarity of the scale items (based on the cosine similarity of their embeddings) to estimate the empirical correlation between these scales (Hommel & Arslan, 2025). As such, the first question one should ask when considering the use of silicon samples is whether or not viable and more robust alternative approaches may be viable instead.

Further, researchers should consider why they would specifically wish to use silicon samples. As described earlier, one of the main reasons for their proposed use is to improve the efficiency and speed of research. However, if (as the work here implies) silicon samples require careful, detailed, and principled decision-making on analytic decisions before they can be used effectively, this begs the question why a researcher would not instead expend their time and effort to simply collect the hard-to-acquire human data. This is particularly salient given that, in all likelihood, the acquisition of at least some human data is required to ensure the silicon sample is performing sufficiently well in the first place (i.e., to conduct the calibration-confirmation approach outlined above). It is, of course, possible that there may be approaches to the development of valid, accurate, and generalisable silicon samples that do not require such data acquisition. However, no such approach currently exists to my knowledge. Researchers should not simply hope that the quality of these samples will improve as new and larger LLMs are developed, given that the observed results across the two studies here did not show any such superiority of newer vs. older LLMs.

Finally, the use of silicon samples merits particular scrutiny and consideration in the context of simulating responses from vulnerable or hard-to-reach populations. If there is not a reliable configuration for accurately simulating data features of the typical research subjects, then it is even less likely that LLMs will be capable of providing accurate insights into the potential responses of individuals from non-WEIRD, vulnerable, and/or hard-to-reach populations. Indeed, the results here show that silicon sample configurations do not generalise in accuracy across even simple data features in typical respondents who are likely very well-represented in these models' training data. The idea that these workflows could faithfully generalise to those individuals who are not well-represented in the training data is even less plausible (Lutz et al., 2025). There is a stark risk that the use of LLMs for this purpose, without careful consideration of these analytic decisions, may create an illusion of representation while in fact speaking over those very same populations (cf. Wang et al., 2025). Unless a more systematic approach to the problem of analytic flexibility is adopted, silicon samples come with the very real risk of hindering the inclusion of these populations in social science, rather than facilitating it.

**Conclusion**

If large language models could be used as valid stand-ins for human participants, this would have a transformative impact on how research in the social sciences is conducted. Silicon samples have the potential to be a major methodological innovation for science, improving the efficiency, accuracy, and representativeness of human research. However, silicon samples could also be yet another tool which produces unreplicable results and leads researchers astray. As things stand, the social sciences have not paid sufficient attention to the many analytic decisions that require careful consideration in the process of generating silicon samples. We face a unique opportunity to get out in front of this issue, and must learn from the lessons of analytic flexibility in the past to prevent yet another unreliable literature. This starts with a recognition that silicon samples should be the outcome of a careful and thoughtful development process which takes nothing for granted.

**Author Note**

Thank you to Malte Elson, Ian Hussey, and Beth Clarke for providing extremely helpful comments on earlier drafts of this manuscript which lead to substantial improvements. All code and data associated with this paper can be found at the Open Science Framework (https://osf.io/skbqm/). Correspondence concerning this manuscript can be sent to Jamie Cummins (jamie.cummins@unibe.ch).